\documentclass[twocolumn,aps,pra,showpacs,amsmath,superscriptaddress]{revtex4}
\usepackage{amsfonts}
\usepackage{amssymb}
\usepackage{mathrsfs}
\usepackage{graphicx}
\usepackage{float}

\begin{document}

\title{Quantum dynamics of hard-core bosons in tilted bichromatic optical lattices}

\author{Xiaoming Cai}
\affiliation{Beijing National
Laboratory for Condensed Matter Physics, Institute of Physics,
Chinese Academy of Sciences, Beijing 100190, China}
\author{Shu Chen}
\affiliation{Beijing National Laboratory for Condensed Matter
Physics, Institute of Physics, Chinese Academy of Sciences, Beijing
100190, China}
\author{Yupeng Wang}
\affiliation{Beijing National Laboratory for Condensed Matter
Physics, Institute of Physics, Chinese Academy of Sciences, Beijing
100190, China}
\date{ \today}

\begin{abstract}
We study the dynamics of strongly repulsive Bose gas in tilted or
driven bichromatic optical lattices. Using the Bose-Fermi mapping
and exact numerical method, we calculate the reduced single-particle
density
matrices, 
and study the 
dynamics of density profile, momentum distribution and condensate
fraction. We show the oscillating and breathing mode of dynamics,
and depletion of condensate for short time dynamics. For long time
dynamics, we clearly show the reconstruction of system at integer
multiples of Bloch-Zener time. We also show how to achieve clear
Bloch oscillation and Landau-Zener tunnelling for many-particle
systems.
\end{abstract}


\maketitle
\section{introduction}
The dynamics of a particle in a period structure has been a
fundamental subject, with the eigenenergies forming the famous Bloch
bands \cite{Bloch1} and eigenstates being delocalized. If a weak
external constant force is introduced, contrary to our intuition,
the particle undergoes an oscillatory motion rather than uniform
motion due to acceleration by the force, which is known as the
famous Bloch oscillation \cite{Bloch1,Dahan1}. Under single-band
tight-binding approximation, eigenenergies of the system form the
Wannier-Stark ladder \cite{Gluck1} with eigenstates localized. Bloch
oscillation bas been observed in semiconductor superlattices for
electrons \cite{Feldmann1}, optical lattices for cold atoms
\cite{Dahan1} and photonic crystals for light pluses
\cite{Pertsch1,Morandotti1}. For a stronger force, a directed motion
is re-introduced by repeated Landau-Zener tunneling to higher Bloch
bands \cite{Landau1,Zener1,Majorana1,Stuckelberg1}. For usual
cosine-shaped potentials, band gaps decrease rapidly as the energy
increases, and this would lead to the decay of Bloch oscillation,
which has been observed in Ref.\cite{Anderson1,Rosam1,Trompeter1}.

In order to study the interplay between Bloch oscillation and
Landau-Zener tunneling, one needs at least a two-band system with
the lowest two bands being well separated from upper ones.
Furthermore, the gap between the lowest two Bloch bands must be
small for observing clear signal of Landau-Zener tunneling. This can
be achieved by bichromatic optical lattices \cite{Bloch2}, where
parameters of the system are adjustable and controllable.
Bichromatic lattices have been implemented by superimposing two
incoherent optical lattices with the wavelength of one lattice being
two times of the other \cite{Gorlitz1,Folling1}, or by virtual
two-photon and four-photon processes
\cite{Ritt1,Salger1,Salger2,Salger3,Kling1}. Under two-band
tight-binding approximation, eigenenergies form two Wannier-Stark
ladders with energy spacing doubled and an offset between them
\cite{Breid1,Breid2}. The corresponding eigenstates are still
localized. The dynamics of a particle is governed by two timescales,
i.e., the Bloch period and the period of Zener oscillation. If the
two periods are commensurate, system reconstructs at integer
multiples of Bloch-Zener time.

So far, most of works concentrate on the dynamics of the
single-particle system. The generalization of these results to
interacting many-particle systems remains an open question. The
possibility to investigate Bloch oscillation and Landau-Zener
tunneling of interacting Bose-Einstein condensate (BEC)
experimentally has attracted much interest. Most of theoretical
studies are based on the mean-field approximation
\cite{Wu1,Zobay1,Liu1,Graefe1} and Gross-Pitaevskii (GP) equations
\cite{Witthaut1}. Results for the dynamics of strongly interacting
many-particle systems are rarely known. In this paper, we study the
dynamics of interacting bosons in bichromatic optical lattices under
constant drag force in the limiting case with infinitely repulsive
interaction which permits us to solve the problem exactly. The
one-dimensional (1D) Bose gas with infinitely repulsive interaction
is known as the hard-core boson (HCB) or Tonks-Girardeau (TG) gas
\cite{Girardeau1}, which can be exactly solved via the Bose-Fermi
mapping \cite{Girardeau1} and has attracted intensive theoretical
attention \cite{Girardeau2,Minguzzi1,Gangardt1}. Experimental access
to the required parameter regime has made the TG gas a physical
reality \cite{Paredes1,Kinoshita1}. Following the exact numerical
approach proposed by Rigol and Muramatsu \cite{Rigol1,Rigol2}, we
calculate the dynamics of density profile, momentum distribution,
and condensate fraction for hard-core bosons in the tilted
bichromatic optical lattice, and show Bloch oscillation,
Landau-Zener tunneling and reconstruction of the system at integer
multiples of Bloch-Zener time.

The paper is organized as follows. In Section II, we present the
model and the exact approach used in this paper. We also recover the
dynamics of single particle system in this section. In Section III,
we study the short-time and long-time dynamics for hard-core bosons
in bichromatic optical lattice with a constant drag force. We also
show how to achieve clear Bloch oscillation and Landau-Zener
tunneling for the many-particle system. Finally, a summary is
presented in Section IV.

\section{model and method}

In the present section we describe the exact approach which we used
to study 1D hard-core bosons in tilted or driven bichromatic optical
lattices. Under the tight-binding approximation, the system can be
described by the following Hamiltonian:
\begin{eqnarray}
\label{eqn1} H=&&-J\sum_i(b^\dagger_i b_{i+1}+
\mathrm{H.c.})+\delta\sum_i(-1)^i n^b_i\notag\\
&&+F\sum_i i n^b_i+\sum_iV_i n^b_i.
\end{eqnarray}
Here we only consider the nearest-neighbor hopping and neglect the
off-diagonal terms of position operator $\hat{x}$ in the Wannier
basis. The operator $b^\dagger_i$ ($b_i$) is the creation
(annihilation) operator of boson which fulfills the hard-core
constraints \cite{Rigol1}, i.e., the on-site anticommutation
($\{b_i, b^\dagger_i\} = 1$) and $[b_i, b^\dagger_j ] = 0$ for
$i\neq j$; $n^b_i$ is the bosonic particle number operator; $J$ is
the hopping amplitude being set to be unit of energy ($J = 1$);
$V_i=V_H(i-i_0)^2$ is the harmonic potential for preparing the
initial state of system, with $V_H$ the strength and $i_0$ the
position of the trap center; $\delta$ is the energy shift of
alternate site; $F$ is the strength of driven force. For
convenience, the lattice spacing is set to be unit.

In order to study the dynamics of the hard-core bosons in driven
optical lattice, we first load the hard-core bosons into a
bichromatic optical lattice with an additional harmonic trap,
then we switch off the harmonic trap and turn on the driven force.
We shall study the evolution of the initially prepared system and
dynamics of the system under the driven force. First, the initial
state is the ground state of Hamiltonian:
\begin{equation}
\label{eqn3} H_\mathrm{init}=-J\sum_i(b^\dagger_i b_{i+1}+
\mathrm{H.c.})+\delta\sum_i(-1)^i n^b_i+\sum_iV_i n^b_i,
\end{equation}
with particle number $N$. In order to get the initial state, it is
convenient to use the Jordan-Wigner transformation \cite{Jordan1}
(JWT):
\begin{equation}
\label{eqn4}
 b^\dagger_j=f^\dagger_j\prod^{j-1}_{\beta=1}e^{-i\pi
f^\dagger_\beta f_\beta},b_j=\prod^{j-1}_{\beta=1}e^{+i\pi
f^\dagger_\beta f_\beta}f_j,
\end{equation}
to map the Hamiltonian of hard-core bosons into the Hamiltonian of
noninteracting spinless fermions $H^{F}_\mathrm{init}$, which is in
the same form as $H_\mathrm{init}$, but with all the boson
operators, e.g.,$b^\dagger_i$, $b_i$ and $n^b_i$, being replaced by
the corresponding fermion operators, e.g., $f^\dagger_i$, $f_i$ and
$n^f_i$. The ground-state wave function of the system with $N$
spinless free fermions, which is a product of lowest N
eigenfunctions, can be obtained by diagonalizing $H^F_\mathrm{init}$
and can be represented as
\begin{equation}
\label{eqn6}
|\Psi^G_F\rangle=\prod^N_{n=1}\sum^L_{i=1}P_{in}f^\dagger_i|0\rangle
,
\end{equation}
where $L$ is the number of lattice sites, $N$ is the number of
fermions (same as bosons), and coefficients $P_{in}$ are the
amplitude of the $n$-th single-particle eigenfunction at $i$-th site
which can form an $L \times N$ matrix $P$ \cite{Cai1}.

After releasing from the trap, the system is described by
Hamiltonian:
\begin{equation}
\label{eqn1} H_{e}=-J\sum_i(b^\dagger_i b_{i+1}+
\mathrm{H.c.})+\delta\sum_i(-1)^i n^b_i+F\sum_i i n^b_i.
\end{equation}
Similar to the above method, from the corresponding free-fermion
Hamiltonian $H^F_e$, we can get all single-particle states and
corresponding energies, and use $P'$ to represent all the
single-particle states, $\varepsilon_i$ to represent the energies.
The nonequilibrium quantum dynamical properties of system can be
calculated through the equal time one-particle Green function which
is defined as:
\begin{equation}
G_{ij}(t)=\langle \Psi_{\mathrm{HCB}}(t)|b_
ib^\dagger_j|\Psi_{\mathrm{HCB}}(t)\rangle,
\end{equation}
with $|\Psi_{\mathrm{HCB}}(t)\rangle$ is the wave function of
hard-core bosons at time $t$ after releasing from harmonic trap.
After some derivations (see Appendix A), one can get
\begin{eqnarray}
G_{ij}(t)=\det \left[ \left( P^{A}\right)^\dagger P^{B}\right],
\end{eqnarray}
where $P^A$ and $P^B$ are obtained from $P$, $P'$ and
$\varepsilon_i$. It follows that the reduced single-particle density
matrix can be determined by the expression
\begin{equation}
\rho_{ij}(t)=\langle
b^\dagger_ib_j\rangle_t=G_{ji}(t)+\delta_{ij}(1-2G_{ii}(t)).
\end{equation}
The momentum distribution is defined by Fourier transform with
respect to $i-j$ of the reduced single-particle density matrix with
the form
\begin{equation}
n(k)=\frac{1}{L}\sum^L_{i,j=1}e^{-ik(i-j)}\rho_{ij}
\end{equation}
where $k$ denotes momentum. The natural orbitals $\phi^\eta_i$ are
defined as eigenfunctions of the reduced single-particle density
matrix \cite{Pensose1},
\begin{equation}
\sum^L_{j=1}\rho_{ij}\phi^\eta_j=\lambda_\eta\phi^\eta_i.
\end{equation}
The natural orbitals can be understood as being effective
single-particle states with occupations $\lambda_\eta$. For
noninteracting bosons, all particles occupy in the lowest natural
orbital and bosons are in the BEC phase at zero temperature, whereas
only the quasicondensation exists for 1D hard-core bosons
\cite{Rigol1}.

For hard-core bosons we know that the state for the corresponding
Fermi Hamiltonian is a product of time-dependent single-particle
states and each single-particle state evolves itself (see
Eq.(\ref{A2}) in Appendix A). We recover the single-particle
properties of the Hamiltonian $H_e$ in Appendix B, which have been
studied by Breid {\it et. al} \cite{Breid1}. From Appendix B, we
know that a single-particle wave function will be reconstructed at
integer multiples of Bloch-Zener time ($T_\mathrm{BZ}$), if $T_1$
and $T_2$ are commensurate. Here $T_1$ is the Bloch period decided
by the spacing of Wannier-Stark ladders, and $T_2$ is the period of
Zener oscillation decided by the offset between Wannier-Stark
ladders. Since each single-particle state can be reconstructed at
period of time, and the period is independent of the single-particle
state and decided by parameters of Hamiltonian, the state for
many-body hard-core bosons composed of a product of single-particle
states will also be reconstructed at integer multiples of
Bloch-Zener time.

\section{quantum dynamics of hard-core bosons}
\begin{figure}[tbp]
\vspace{2.2cm} \hspace{-2.2cm}
\includegraphics[width=5.5cm, height=4.5cm, bb=25 20 303 235]{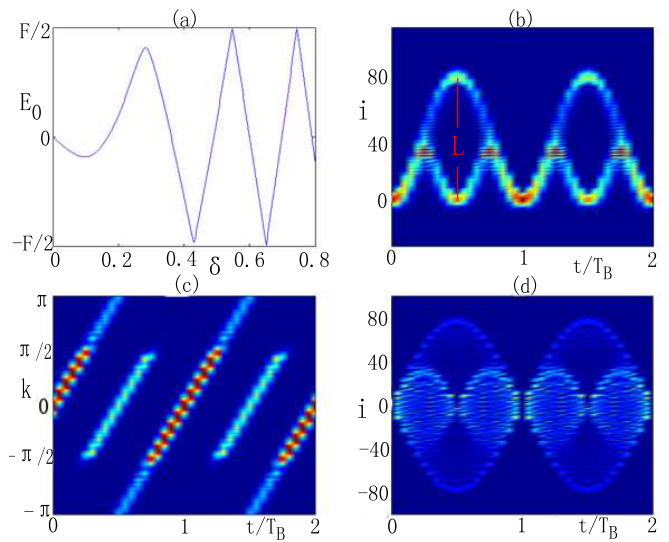}
\caption{(Color online) (a): $E_0$ vs $\delta$ for the system with
$F=0.05$. The dynamics of density profile (b) and momentum
distribution(c) for single particle system with $\delta=0.1684$,
$F=0.05$, and $V_H=0.001$. (d): The breathing mode dynamics of
density profile for single particle system with $\delta=0.1684$,
$F=0.05$, and $V_H=20$.} \label{Fig1}
\end{figure}

In order to observe reconstruction of system, two periods $T_1$ and
$T_2$ must be commensurate which are decided by $F$ and $E_0$, where
$E_0$ is a function of $F$ and $\delta$. In Fig.\ref{Fig1}a, we show
numerical results of $E_0$ versus $\delta$ for a particular $F$. For
different $F$, structure of the picture is similar. In order to
generate particular Bloch-Zener time ($T_\mathrm{BZ}$), $\delta$
must be one of the discrete numbers. For example, if we want
$T_\mathrm{BZ}=T_B$ for the system with $F=0.05$, we have to let
$E_0=0$ and then $\delta=0.1684,0.3614,...$.

For comparison, we first recover the dynamics of a single particle
system. We choose $E_0=0$ and let system reconstruct at integer
multiples of Bloch time ($T_B$). From now on, we take Bloch time
$T_B=2T_1$ as the reference timescale. In Fig.\ref{Fig1}b, we show
the dynamical evolution of density profile for a single-particle
system, from which one can see the reconstruction of the density
profile at integer multiples of Bloch time. The edge of Brillouin
zone is reached at $t=T_B/4$, and part of particle moves into upper
excited Bloch band located in upper half of the figure. The particle
in upper excited band returns to lower Bloch band at $t=3T_B/4$. For
quantitative analysis of Landau-Zener tunneling rate, one can
characterize it by the number of particles in upper half of picture
at time $t=T_B/2$. Numerical results show that the Landau-Zener
tunnelling probability \cite{Witthaut1}:
\begin{eqnarray}
P_\mathrm{LZ}\approx \mathrm{exp}(-\frac{\pi\delta^2}{2F}).
\end{eqnarray}
So in order to see a clear signal of Landau-Zener tunneling, we have
to choose small $\delta$ for a given strength of force $F$.
Furthermore, the available interval for motion of a particle in
lattice \cite{Hartmann1}:
\begin{eqnarray}
L\approx 4/F.
\end{eqnarray}
In Fig.\ref{Fig1}c, we also show the dynamical evolution of momentum
distribution of the single-particle system. Particles with momentum
in the interval $(-\pi/2,\pi/2)$ are in lower Bloch band and outside
the region particles are in upper excited Bloch band. From this
picture we can see the clear signals for Bloch oscillation and
Landau-Zener tunneling. Furthermore the momentum is linear with time
with slope given by $F$. Next we consider a localized initial state
which has a wide momentum distribution and can be implemented by
setting a strong harmonic trap potential, for example, $V_H=20$
here. From Fig.\ref{Fig1}d, one can find a breathing behavior of
density profile. The reconstruction still happens at integer
multiples of $T_B$. But one can observe that the enveloping
structure with period of $T_B$ is overlayed by a breathing mode of
smaller amplitude, whereas part of particle remains in lower Bloch
band all the time with period of $T_B/2$.

\begin{figure*}[tbp]
\vspace{3.5cm} \hspace{-8.2  cm}
\includegraphics[width=8.4cm, height=6cm, bb=25 20 303
235]{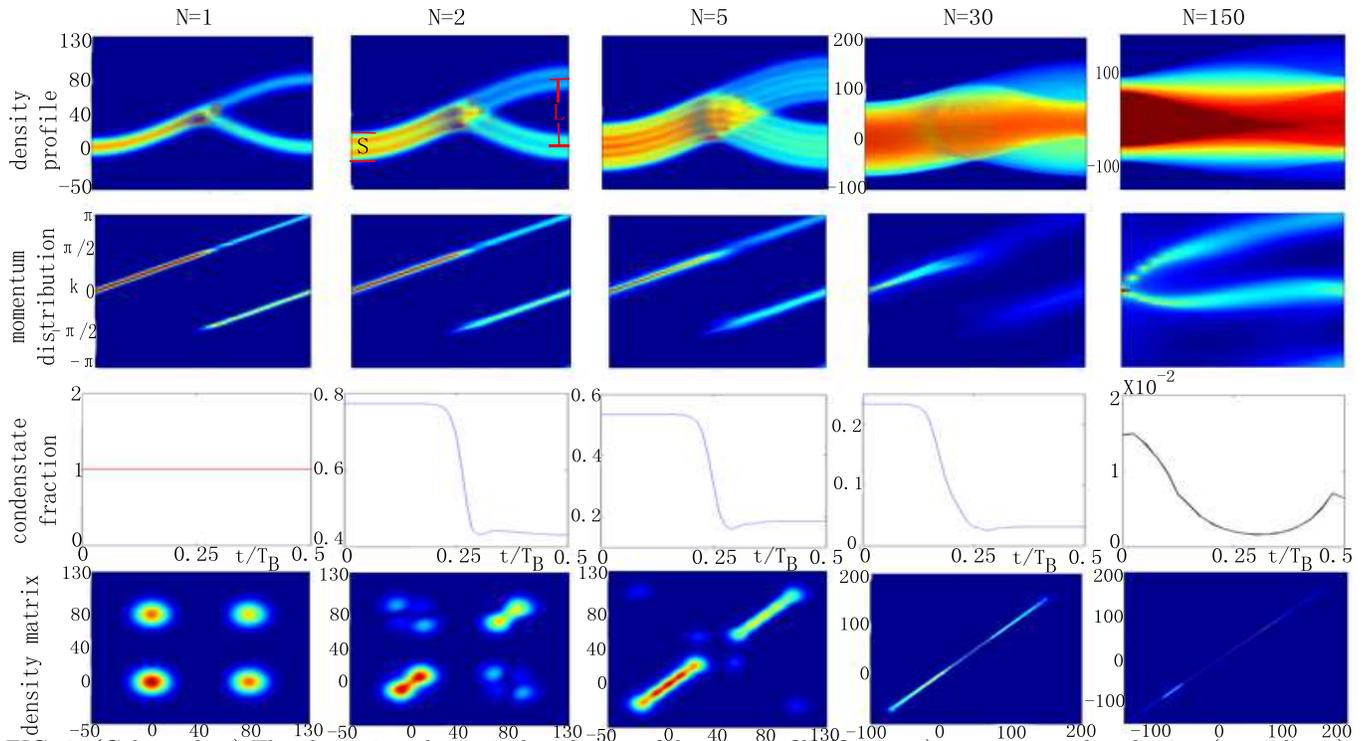} \caption{(Color online) The short-time dynamical
evolution of density profile (first row), momentum distribution
(second row), condensate fraction (third row), and modules of the
reduced single-particle density matrix (fourth row) at time
$t=T_B/2$ for systems of different particle number with $F=0.05$,
$\delta=0.1684$, $E_0=0$, and $V_H=10^{-4}$ ($V_H=8\times10^{-4}$
for the fifth column).} \label{Fig2}
\end{figure*}

Now we consider the case of many-body hard-core bosons and study the
short time dynamics firstly. Systems with various particle numbers
will be considered by keeping the other parameters fixed. For
comparison, we show the dynamics of a single-particle system in the
first column of Fig.\ref{Fig2}. After releasing from trap, the
particle speeds up under the drag force $F$, and it reaches the edge
of Brillouin zone at time $T_B/4$. Part of particle moves into upper
excited Bloch band through Landau-Zener tunneling, while the other
part of particle remains in lower Bloch band with changing of sign
of momentum by Brag scattering. Because of Landau-Zener tunneling,
the particle turns into two parts and they are separated in the real
space with particles in the upper half of the density distribution
being in the upper excited Bloch band. In momentum distribution, the
particles outside the first Brillouin zone of ($-\pi/2,\pi/2$) are
in upper excited Bloch band. The changing of sign of momentum occurs
at time $T_B/4$ by Brag scattering for particles in the lower Bloch
band. In the third row of Fig.\ref{Fig2}, we show the evolution of
condensate fraction which is defined as $\lambda_0/N$ with
$\lambda_0$ being the occupation of the lowest effective
single-particle state. For single particle system, the particle is
always in the lowest effective single-particle state, and the
condensate fraction is one all the time.  For the 1D many-body
systems of hard-core bosons, there are only quasi-condensation with
$\lambda_0 \propto \sqrt{N}$ \cite{Girardeau1,Rigol1}. In the fourth
row of Fig.\ref{Fig2}, we show the reduced single-particle density
matrix at time $T_B/2$. Here we consider the modulus because of the
elements of density matrix being complex numbers after turning off
trap. The upper right spot in picture is caused by particles in
upper excited Bloch band and the lower left spot for lower Bloch
band. For single particle dynamics, however particles in upper
excited Bloch band and lower Bloch band are separated in real space,
there are phase coherence between them, the off-diagonal parts of
reduced single-particle density matrix are very strong.

In the second to fifth column, we show the dynamics for the
hard-core bosons with $N=2,5,30$ and $150$, respectively. As
particle number increases, the adding particle has to occupy higher
single-particle state because one state can only be occupied by a
hard-core boson.
And the size ($S$) of system becomes larger and larger, while the
available interval $L$ of the system decided by force $F$ remains
unchanged. Also, the width of momentum distribution becomes wider
for the larger system. As the momentum distribution becomes wider,
it takes shorter time for particles at the edge of the momentum
distribution to reach the edge of Brillouin zone, and thus
Landau-Zener tunneling happens early, which leads to the condensate
fraction decreasing early. The condensate fraction decreases
($\propto 1/\sqrt{N}$) as the particle number increases. As time
increases but is smaller than $T_B/4$, the condensate fraction
basically does not change. A slight increase of the condensate
fraction in short time is caused by the expanding after turning off
trap \cite{Cai1}. At time $t=T_B/4$, Landau-Zener tunneling happens,
and the condensate fraction decreases quickly. After this there
appears an overdamped area, and then the condensate fraction keeps
unchanged basically. For the reduced single-particle density matrix
at time $T_B/2$, lengths for the two parts of diagonal terms become
larger as the size ($S$) of system increases. The off-diagonal parts
of matrix become weaker as particles add in. Also, the phase
coherence between particles in the upper excited Bloch band and
lower Bloch band decreases. As shown in the figure, the particles
between upper band and lower band lost their phase coherence when
$N=30$. Furthermore, the particles in the upper excited Bloch band
and lower Bloch band developed phase coherence inside each part, the
reduced single-particle density matrix has exponential-law decay in
each part as the distance increases.

We note that the density profile no longer splits into two obviously
separated parts after time $T_B/4$ for system with $N=30$ as shown
in the fourth column, where the particles in the upper excited Bloch
band and lower Bloch band are overlapped in real space. To achieve
this situation, one has to adjust the parameters of system and let
the size of system larger than the available interval ($S>L$). Also
one has to avoid  the initial state staying in a localized state
because localized initial state will cause the breathing-mode
dynamics. Although the two parts are overlapped in real space, they
are separated in the momentum distribution. The width of the
momentum distribution becomes larger as the particle number
increases. The dynamical evolution of condensate fraction still has
similar structure, except that the quick decrease happens more
early.
As two parts of particles are overlapped in real space, the diagonal
terms of density matrix are also overlapped. For sites outside the
overlapped area in upper excited Bloch band or lower one, the
density matrix still has a exponential-law decay as distance
increases. For overlapped area, the density matrix is irregular, but
as the distance increases, it goes to zero quickly. In the fifth
column, we show the dynamics of system with $N=150$. As the particle
number increases, the localization of initial state becomes stronger
\cite{Rigol1}, and we see the breathing mode of dynamics of density
distribution. Lots of particles are localized in the center, and
particles in two parts are overlapped. The dynamics of momentum
distribution is also changed. The momentum distribution does not
increase linearly as time increases, and momentum distributions for
particles in the upper excited Bloch band and lower Bloch band are
overlapped. For localized initial state, the momentum distribution
is almost flat. Right after turning off trap, there are particles
moving into upper excited Bloch band through Landau-Zener
tunnelling, and the condensate fraction decreases immediately.
Meanwhile, for a localized initial state, the expanding is also
important after turning off trap, and there is a large overdamped
area in the breathing mode dynamics of condensate fraction. The
reduced single-particle density matrix still has exponential-law
decay for particles in localized states.

\begin{figure}[tbp]
\vspace{6.1cm} \hspace{-2.5cm}
\includegraphics[width=5.5cm, height=5cm, bb=25 20 303 235]{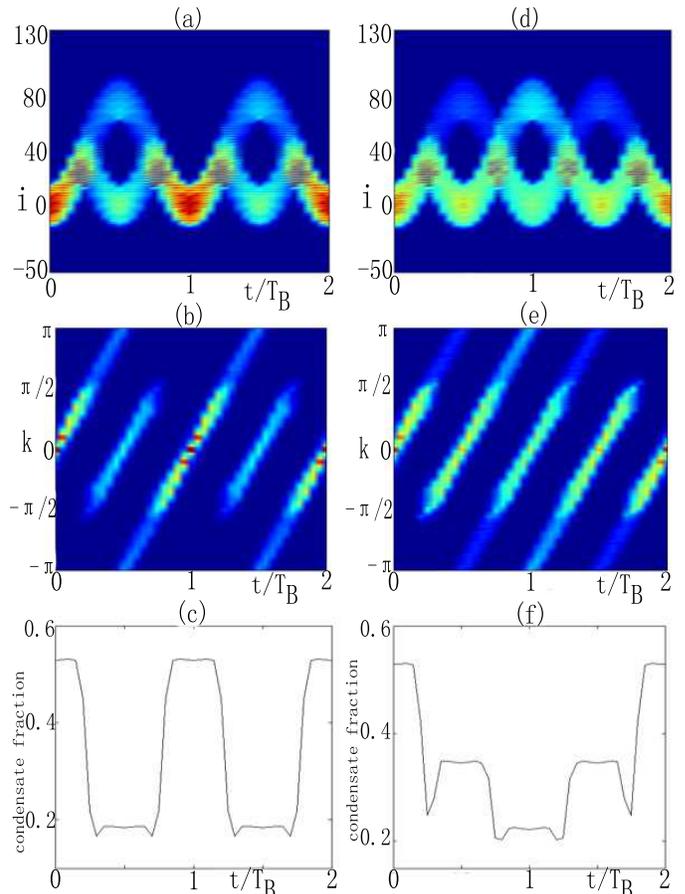}
\caption{(Color online) The dynamical evolution of density profile
(a,d), momentum distribution (b,e), and condensate fraction (c,f)
for five-particle system with $F=0.05$, $V_H=10^{-3}$ and
$\delta=0.1684$ (a,b,c), $\delta=0.238$ (d,e,f).} \label{Fig3}
\end{figure}

Next, we consider the long-time dynamics of hard-core bosons.
In Fig.\ref{Fig3}a, we show the dynamical evolution of density
distribution for five-particle system with Bloch-Zener time
$T_\mathrm{BZ}=T_B$, where $E_0=0$. After releasing from the
harmonic trap, particles move along with the direction of the force
$F$. Around time $t=T_B/4$, particles reach the edge of Brillouin
zone, and part of particles move into the upper excited Bloch band
through Landau-Zener tunneling and keep moving along with force $F$.
The other part of particles remain in the lower Bloch band, but the
momentum of particles changes the sign due to Brag scattering and
thus the particles move against the force $F$. Around time
$t=3T_B/4$, particles reach the edge of Brillouin zone again, and
all particles move into the lower Bloch band. At time $t=T_B$, the
system reconstructs into the initial state. As time goes on, more
periods occur. The dynamical evolution of the momentum distribution
for the same system is shown in Fig.\ref{Fig3}b. After turning off
trap, particles accelerate under force $F$, and the momentum
increases linearly with time. Around time $t=T_B/4$, particles reach
the edge of Brillouin zone ($k=\pi/2$). Part of particles, which
move into the upper Bloch band through Landau-Zener tunnelling,
remain in the same belt and their momentum increases linearly with
time. The other part of particles keep in the lower Bloch band and
change the sign of their momentum by Brag scattering. Two parts of
particles are recombined around time $t=3T_B/4$, and the momentum
distribution returns to the initial distribution at $t=T_B$. In
Fig.\ref{Fig3}c, we show the dynamical evolution of condensate
fraction. Around time $t=T_B/4$,  the condensate fraction has a
quick decrease due to Landau-Zener tunneling. After this there is an
overdamped area. Around time $t=3T_B/4$, two parts of particles
recombine, and the condensate fraction increases quickly. The system
returns to its initial state at time $T=T_\mathrm{BZ}$. Furthermore,
the condensate fraction is symmetrical with center at
$T=T_\mathrm{BZ}/2$. In Fig.\ref{Fig3}, we also show the dynamics of
system with $T_\mathrm{BZ}=2T_B$ where $E_0=F/4$, which has similar
properties with the previous one. From Fig.\ref{Fig3}d, one can
obviously observe the reconstruction of system after Bloch-Zener
time $T_\mathrm{BZ}$, Bloch oscillation and Landau-Zener tunneling.

\begin{figure}[tbp]
\vspace{6.1cm} \hspace{-2.5cm}
\includegraphics[width=5.5cm, height=5cm, bb=25 20 303 235]{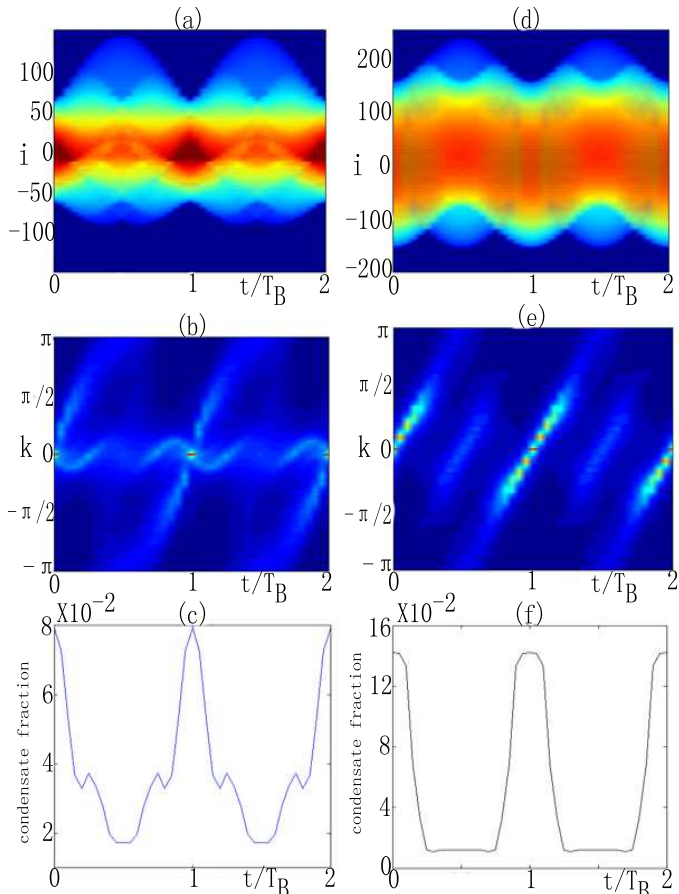}
\caption{(Color online) The dynamical evolution of density profile
(a,d), momentum distribution (b,e), and condensate fraction (c,f)
for the eighty-particle system with $F=0.05$, $\delta=0.1684$ and
$V_H=10^{-3}$ (a,b,c), $V_H=4\times10^{-5}$ (d,e,f).} \label{Fig4}
\end{figure}

In Fig.\ref{Fig4}, we show the dynamical evolution of system with
eighty hard-core bosons. As particles add in, the particles in the
initial state get more localized, and the dynamical evolution of the
system is dominated by the breathing mode instead of the oscillating
mode. In Fig.\ref{Fig4}a, we show the dynamical evolution of the
density profile for system with $F=0.05$, $\delta=0.1684$ and
$V_H=10^{-3}$. This picture is not in prefect breathing mode because
of the initial state being not localized enough. First of all, the
reconstruction of density profile happens again after integer
multiples of Bloch-Zener time $T_\mathrm{BZ}$. Second, lots of
particles are localized at the center and form a belt. Third, one
can observe that the enveloping structure is overlayed by a
breathing mode of smaller amplitude. The outside breathing mode is
formed by Bloch oscillation and Landau-Zener tunneling for particles
in the upper excited Bloch band with period $T_\mathrm{BZ}=T_B$. The
inside breathing mode is formed by Bloch oscillation for particles
remaining in the lower Bloch band with period $T_1=T_B/2$. Center is
the localized belt. In Fig.\ref{Fig4}b we show the dynamical
evolution of the momentum distribution. Strip structure disappears
and windmill structure appears with the center at $(nT_B,k=0)(n\in
\mathbb{N})$. In this picture we can not distinguish the two parts
of particles, and lots of particles are localized at area of
$k\approx 0$ all the time. Foremost, the momentum distribution
reconstructs at integer multiples of $T_\mathrm{BZ}$. The dynamical
evolution of the condensate fraction is shown in Fig.\ref{Fig4}c, As
the initial state is a localized state, there are many particles
with high momentum, and right after turning off trap the condensate
fraction decreases quickly, and it reconstructs at
$t=T_\mathrm{BZ}$. As a lot of particles add in, the initial state
becomes a localized state and one can observe breathing mode
dynamical evolution of density profile. In order to observe the
oscillating mode and Landau-Zener tunneling, we have to reduce the
localization of initial state by decreasing the strength of the
harmonic trap. The dynamics of system after decreasing the strength
of trap is shown in the second column of Fig.\ref{Fig4}. For the
density distribution we actually see the oscillating mode, but we
can not see the Landau-Zener tunneling clearly. At time range
$(T_B/4,3T_B/4)$, the particles in the upper excited Bloch band and
lower Bloch band are overlapped in real space. This is due to the
available interval for the motion being $L\approx4/F=80$ for system
with $F=0.05$, but the size of system is about $S\approx 300$. It is
clear that $S>L$ and the overlapped structure appears. After
decreasing the strength of trap, the strip structure reappears in
the dynamical evolution of the momentum distribution instead of the
windmill structure. Furthermore the momentum distribution between
the particles in the upper excited and lower Bloch band are
distinguishable despite they are overlapped in real space. For the
condensate fraction, the curve is flat again for the time short
after turning off trap.

\begin{figure}[tbp]
\vspace{-1.3cm} \hspace{-2.7cm}
\includegraphics[width=5.5cm, height=5cm, bb=25 20 303 235]{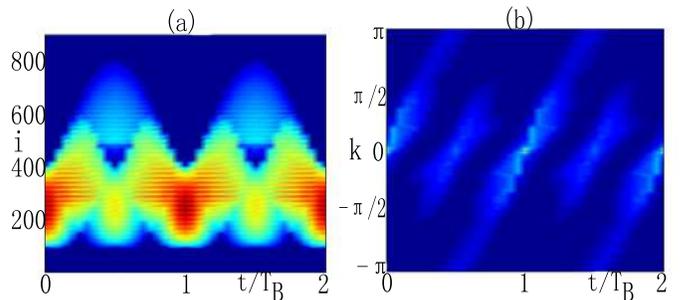}
\caption{(Color online) The dynamical evolution of density profile
(a), momentum distribution (b), for the eighty-particle system with
$F=0.01$, $\delta=0.08211$ and $V_H=4\times10^{-5}$.} \label{Fig5}
\end{figure}

So in order to observe the oscillating mode of dynamics and
Landau-Zener tunneling, one has to let $L>S$. Now we have to
decrease the strength of force $F$ to achieve a bigger available
interval for the motion. Once $F$ is changed , we have to change
$\delta$ too to make $E_0=0$ for the system still with Bloch-Zener
time $T_\mathrm{BZ}=T_B$. Furthermore we have to choose a small
$\delta$ to achieve the big enough Landau-Zener tunneling
probability, otherwise we only see the Bloch oscillation but can not
see the Landau-Zener tunneling. In Fig.\ref{Fig5}a, we show the
dynamical evolution of density profile for the system with $F=0.01$,
$\delta=0.08211$ and $T_\mathrm{BZ}=T_B$. Now we see the clear
signals of Bloch oscillation and Landau-Zener tunneling. The
dynamical evolution of the momentum distribution and condensate
fraction are similar to Fig.\ref{Fig4}e and Fig.\ref{Fig4}f.

\section{conclusion}

In summary, we have studied the dynamics of infinitely repulsive
Bose gas in tilted or driven bichromatic optical lattices. Using the
Bose-Fermi mapping and exact numerical method, we calculate the
one-particle density matrices, density profiles, momentum
distributions, natural orbitals and their occupations (condensate
fraction). Both the short-time and long-time dynamical evolution of
density profile, momentum distribution and condensate fraction are
studied. It is clearly shown that the reconstruction of system at
integer multiples of Landau-Zener time. We also give estimations for
how to achieve clear Bloch oscillation and Landau-Zener tunneling in
given many-particle systems.

\begin{acknowledgments}
This work has been supported by NSF of China under Grants
No.10821403 and No.10974234, programs of Chinese Academy of Science,
and National Program for Basic Research of MOST.
\end{acknowledgments}

\begin{appendix}

\section{THE EQUAL TIME GREEN FUNCTION FOR HARD-CORE BOSONS}
The equal time Green function for the hard-core bosons can be
written in the form
\begin{eqnarray}
G_{ij}(t)&=&\langle\Psi_{\mathrm{HCB}}(t)|b_ib^\dagger_j|\Psi_{\mathrm{HCB}}(t)\rangle
\notag\\
 &=& \langle\Psi_F(t)|\prod_{\beta=1}^{i-1}e^{i\pi
f^\dagger_\beta f_\beta}f_if^\dagger_j\prod_{\gamma=1}^{j-1}e^{-i\pi
f^\dagger_\gamma f_\gamma}|\Psi_F(t)\rangle \notag\\
&=&\langle\Psi^B|\Psi^A\rangle
\end{eqnarray}
where $|\Psi_{\mathrm{HCB}}(t)\rangle$ is the wave function of
hard-core bosons at time $t$ after releasing from harmonic trap and
$|\Psi_F(t)\rangle$ is the corresponding one for noninteracting
fermions. In addition, we denote
\begin{eqnarray}
\left|
\Psi^A\right\rangle&=&f^\dagger_j\prod_{\gamma=1}^{j-1}e^{-i\pi
f^\dagger_\gamma f_\gamma}|\Psi_F(t)\rangle,\notag\\
\left|\Psi^B\right\rangle&=&f^\dagger_i\prod_{\beta=1}^{i-1}e^{-i\pi
f^\dagger_\beta f_\beta}|\Psi_F(t)\rangle.
\end{eqnarray}

The wave function $|\Psi_F(t)\rangle$ can be easily calculated with
the initial wave function $|\Psi^G_F\rangle$
\begin{eqnarray}
\label{A1}
|\Psi_F(t)\rangle=e^{-iH^F_et}|\Psi^G_F\rangle=\prod_{n=1}^{N}\sum_{l=1}^LP_{ln}(t)f_l^{\dagger
}|0\rangle,
\end{eqnarray}
with
\begin{eqnarray}
\label{A2}
P_{ln}(t)=\sum^N_{k=1}e^{-i\varepsilon_kt}P'_{lk}\sum^N_{j=1}(P'^*_{jk}P_{jn}),
\end{eqnarray}
where we have set $\hbar=1$ in evolution operator, and $P(t)$ is the
matrix of $|\Psi_F(t)\rangle$ in the same way as $|\Psi^G_F\rangle$.
In order to get Eq.(\ref{A1}), one has to insert
$\sum_{j=1}^L|\phi_j\rangle\langle\phi_j|=1$ in to it, where
$|\phi_j\rangle=\sum_{n=1}^LP'_{nj}f^\dagger_n|0\rangle$ is the
lowest j-th eigenfunction of $H^F_e$.We can see that
$|\Psi_F(t)\rangle$ is still a product of time-dependent
single-particle states.

In order to calculate $\Psi^A$(and $\Psi^B$) we notice that
\begin{eqnarray}
\prod_{\gamma=1}^{j-1}e^{-i\pi f^\dagger_\gamma
f_\gamma}=\prod_{\gamma=1}^{j-1}[1-2f^\dagger_\gamma f_\gamma].
\end{eqnarray}
Then, the action of $\prod_{\gamma=1}^{j-1}e^{-i\pi f^\dagger_\gamma
f_\gamma}$ on the state $|\Psi_F(t)\rangle$ (Eq.(\ref{A1}))
generates only a change of sign on the element $P_{ln}(t)$ for
$l<j$, and one has to add a column to $P(t)$ with element
$P_{j,N+1}=1$ and all the others equal to zero for the further
creation of a particle at site j. Then
\begin{eqnarray}
\left|
\Psi^A\right\rangle&=&\prod_{n=1}^{N+1}\sum_{l=1}^LP_{ln}^{A}f_l^{\dagger
}\left| 0\right\rangle\notag\\
\left|\Psi^B\right\rangle&=&\prod_{n=1}^{N+1}\sum_{l=1}^LP_{ln}^{B}f_l^{\dagger
}\left| 0\right\rangle
\end{eqnarray}
where $P^A$ and $P^B$ are obtained from $P(t)$ changing the required
signs and adding the new column.

The Green function is written as
\begin{eqnarray}
G_{ij}(t)&=&\langle0|\prod_{n=1}^{N+1}\sum_{l=1}^LP^{B*}_{ln}f_l\prod_{n'=1}^{N+1}\sum_{l'=1}^LP^{A}_{ln}f_l'|0\rangle\notag\\
&=&\sum_{l_1\cdot\cdot\cdot1_{N+1},l'_1\cdot\cdot\cdot1'_{N+1}}^{L}P^{B*}_{l_11}\cdot\cdot\cdot
P^{B*}_{l_{N+1}N+1}\notag\\
&&\times P^{A}_{l'_11}\cdot\cdot\cdot
P^{A}_{l'_{N+1}N+1}\langle0|f_{l_1}\cdot\cdot\cdot
f_{l_{N+1}}f^\dagger_{l'_{l_{N+1}}}\cdot\cdot\cdot
f^\dagger_{l'_1}|0\rangle\notag\\
&=&\mathrm{det}[(P^B)^\dagger P^A]
\end{eqnarray}
which requires
\begin{eqnarray}
\langle0|f_{l_1}\cdot\cdot\cdot
&&f_{l_{N+1}}f^\dagger_{l'_{l_{N+1}}}\cdot\cdot\cdot
f^\dagger_{l'_1}
|0\rangle=\notag\\
&&\varepsilon^{\lambda_1\cdot\cdot\cdot\lambda_{N+1}}\delta_{l_1l'_{\lambda_1}}\cdot\cdot\cdot\delta_{l_{N+1}l'_{\lambda_{N+1}}}
\end{eqnarray} with
$\varepsilon^{\lambda_1\cdot\cdot\cdot\lambda_{N+1}}$ the
Levi-Civita symbol and $\lambda=1\cdot\cdot\cdot N+1$.

\section{THE SINGLE-PARTICLE PROPERTIES OF HAMILTONIAN $H_e$}

First of all, for the field-free case with $F=0$, a straightforward
calculation yields the dispersion relation
\begin{equation}
E_{\beta k}=(-1)^{\beta+1}\sqrt{\delta^2+4\mathrm{cos}^2(k)},
\end{equation}
and corresponding wavefunctions $|\chi_\beta(k)\rangle$ (Bloch bands
and Bloch waves) with the miniband index $\beta=0,1$. For nonzero F,
the spectrum of Hamiltonian consists of two Wannier-Stark ladders
with an offset in between. After introducing translation operator
\begin{equation}
T_m=\sum^\infty_{n=-\infty}b^\dagger_{n-m}b_n
\end{equation}
and an operator $G$ that causes the inversion of sign of $\delta$ in
Hamiltonian:
\begin{equation}
GH_e(\delta)=H_e(-\delta)G,\quad\quad[T_m,G]=0,
\end{equation}
an eigenvector $|\Psi\rangle$ of $H_e$ with the eigenvalue
$E(\delta,F)$ satisfies:
\begin{eqnarray}\label{eqn7}
&&H_e\{T_{2l}|\Psi\rangle\}=\{E(\delta,F)-2lF\}\{T_{2l}|\Psi\rangle\}\\
&&H_e\{T_{2l+1}|\Psi\rangle\}=\{E(-\delta,F)-(2l+1)F\}\{T_{2l+1}|\Psi\rangle\}.\notag
\end{eqnarray}
Thus, the eigenenergies of Hamiltonian
\begin{eqnarray}
E_{0,n} &=& E(\delta,F)+2nF\notag\\
E_{1,n} &=& E(-\delta,F)+(2n+1)F, \label{eqn9}
\end{eqnarray}
consists of two Wannier-Stark ladders with the corresponding
eigenstates
$|\Psi_{\beta,n}\rangle=T_{-(2n+\beta)}G^\beta|\Psi\rangle$
\cite{Breid1}. A further calculation can prove that $E_0\equiv
E(\delta,F)=-E(-\delta,F)$.

For an initial state expanded in Wannier-Stark basis:
\begin{eqnarray}
|\Phi\rangle=\sum_n c_{0,n}|\Psi_{0,n}\rangle+\sum_n
c_{1,n}|\Psi_{1,n}\rangle,
\end{eqnarray}
the dynamics of $|\Phi\rangle$ under Hamiltonian $H_e$ is given by
\begin{eqnarray}
|\Phi(t)\rangle=\sum_n
c_{0,n}e^{-iE_{0,n}t}|\Psi_{0,n}\rangle+\sum_n
c_{1,n}e^{-iE_{1,n}t}|\Psi_{1,n}\rangle.&&\notag\\
&&
\end{eqnarray}
Expanding Wannier-Stark functions in Bloch basis:
\begin{eqnarray}
|\Psi_{\beta,n}\rangle=\int^{\tfrac{\pi}{2}}_{-\tfrac{\pi}{2}}a_{\beta
n}(k)|\chi_0(k)\rangle\mathrm{d}k+\int^{\tfrac{\pi}{2}}_{-\tfrac{\pi}{2}}b_{\beta
n}(k)|\chi_1(k)\rangle\mathrm{d}k,&&\notag\\
&&
\end{eqnarray}
and projecting $|\Phi(t)\rangle$ onto Bloch basis, one can get
\begin{widetext}
\begin{eqnarray}
\label{eqn8}
&&\langle\chi_0(k)|\Phi(t)\rangle=e^{-iE_0t}[a_{0,0}(k)C_0(k+Ft)+a_{1,0}(k)e^{-i(F-2E_0)t}C_1(k+Ft)],\notag\\
&&\langle\chi_1(k)|\Phi(t)\rangle=e^{-iE_0t}[b_{0,0}(k)C_0(k+Ft)+b_{1,0}(k)e^{-i(F-2E_0)t}C_1(k+Ft)],
\end{eqnarray}
\end{widetext}
where $C_\beta$ are the Fourier series of $c_{\beta,n}$:
\begin{eqnarray}
C_\beta(k+Ft)=\sum_nc_{\beta,n}e^{-i2n(k+Ft)},
\end{eqnarray}
which are $\pi$-periodic. To get Eq.(\ref{eqn8}) one has to use
$T_{-2n}|\chi_\beta(k)\rangle=e^{-i2nk}|\chi_\beta(k)\rangle$
(translation of Bloch waves). From Eq.(\ref{eqn8}), one can see that
the dynamics of a particle is characterized by two periods:
$C_\beta$ are functions with period of
\begin{eqnarray}
T_1=\frac{\pi}{F},
\end{eqnarray}
whereas the exponential function $e^{-i(F-2E_0)t}$ has a period of
\begin{eqnarray}
T_2=\frac{2\pi}{F-2E_0}.
\end{eqnarray}
$T_1$ is half of the Bloch time $T_B=2\pi/F$ for the single band
system($\delta=0$). In general if $T_1$ and $T_2$ are commensurate,
\begin{eqnarray}
\frac{T_1}{T_2}=\frac{2F}{F-2E_0}=\frac{m}{n} \quad\mathrm{with
}\quad n,m\in \mathbb{N},
\end{eqnarray}
thus the wavefunction is reconstructed at integer multiples of
Bloch-Zener time ($T_\mathrm{BZ}=nT_1$).

\end{appendix}

\end{document}